\documentclass[showpacs,preprintnumbers,amssymb,twocolumn,aps,reprint,superscriptaddress,showkeys,amsmath,floatfix]{revtex4-2}

\usepackage{graphicx}
\usepackage{dcolumn}
\usepackage[dvipsnames]{xcolor}
\usepackage{bm}
\usepackage{amsmath}
\usepackage{amssymb}
\usepackage{epsfig}
\usepackage{amsfonts}
\usepackage{lineno,hyperref}
\usepackage{array}
\usepackage{float}
\usepackage{microtype}
\usepackage{multirow}
\usepackage{adjustbox}
\usepackage[english]{babel}
\usepackage{epstopdf}
\usepackage{blindtext}
\usepackage{booktabs}
\usepackage{subcaption}
\usepackage[a4paper, total={6.5in, 10in}]{geometry}
\usepackage{appendix}

\def \a{\alpha}

\def \L{\Lambda}

\def \g{\gamma}

\def \O{\Omega}

\def \t{\theta}

\def \be{\begin{equation}}
\def \ee{\end{equation}}
\def \ben{\begin{eqnarray}}
\def \een{\end{eqnarray}}

\def \n{\nonumber}
\def \G{\bar{G}}

\def \La{\mathcal{L}}

\def \t{\Tilde}

\begin{document}
\title{
Non-Affine Extensions of the Raychaudhuri Equation in the K-essence Framework }

\author{Samit Ganguly}
\email{samitgphy07@gmail.com}
\affiliation{Department of Physics, University of Calcutta, 92, A.P.C. Road, Kolkata-700009, India}
\affiliation{Department of Physics, Haldia Government College, Haldia, Purba Medinipur 721657, India}

\author{Goutam Manna$^a$}
\email{goutammanna.pkc@gmail.com \\$^a$Corresponding author}
\affiliation{Department of Physics, Prabhat Kumar College, Contai, Purba Medinipur 721404, India} 
\affiliation{Institute of Astronomy Space and Earth Science, Kolkata 700054, India}

\author{Debashis Gangopadhyay}
\email{debashis.g@snuniv.ac.in}
\affiliation{Department of Physics, School of Natural Sciences, Sister Nivedita University, DG 1/2, Action Area 1 Newtown, Kolkata 700156, India.}

\author{Eduardo Guendelman}
\email{guendel@bgu.ac.il}
\affiliation{Department of Physics, Ben-Gurion University of the Negev, Beer-Sheva, Israel}
\affiliation{Frankfurt Institute for Advanced Studies (FIAS), Ruth-Moufang-Strasse 1, 60438 Frankfurt am Main, Germany.}
\affiliation{Bahamas Advanced Study Institute and Conferences, 4A Ocean Heights, Hill View Circle, Stella Maris, Long Island, The Bahamas.}

\author{Abhijit Bhattacharyya}
\email{abhattacharyyacu@gmail.com}
\affiliation{Department of Physics, University of Calcutta, 92, A.P.C. Road, Kolkata-700009, India}

\date{\today}

\begin{abstract}
We present a new avenue of the Raychaudhuri Equation (RE) by introducing a non-affine parametrization within the k-essence framework. This modification accounts for non-geodesic flow curves, leading to emergent repulsive effects in cosmic evolution. Using a DBI-type k-essence Lagrangian, we derive a modified RE and demonstrate its ability to address the Hubble tension while predicting a natural emergence of a dynamical dark energy equation of state. Our Bayesian analysis, constrained by cosmological data, supports the theoretical scaling relation of the k-essence field ($\dot{\phi}$) and the cosmic scale factor ($a$). Furthermore, we reinterpret the modified RE as an anti-damped harmonic oscillator, we found a caustic avoidance signature, it may reveal classical or quantum-like effects in cosmic expansion. These results suggest a deep connection between scalar field dynamics and modified gravity, offering new perspectives on the nature of the expansion history of the universe.
\end{abstract}



\maketitle




\section{Introduction} 
The Raychaudhuri equation (RE) \cite{Raychaudhury1, Raychaudhury2}, a fundamental geometric identity formulated as a frame-independent scalar equation \cite{Kar, Dadhich, Hensh} serves as a cornerstone of general relativity. The RE is formulated in a purely geometric way without using any other gravity theory. It governs the evolution of geodesic congruences and plays a crucial role in understanding gravitational collapse, singularity formation, and cosmic expansion \cite{Penrose, Hawking1, Hawking2}. Traditionally, RE is formulated for affinely parameterized geodesics, ensuring geodesic flow under purely gravitational interactions. However, recent advancements suggest that non-affine parametrization can encode additional physical effects such as non-trivial interactions and emergent repulsive forces \cite{Kar, Dadhich, Hensh}.

While Raychaudhuri’s original equation is formulated in a general form applicable to both affine and non-affine parametrization \cite{Kar, Dadhich}, the non-affine case remains largely unexplored in the literature. In this work, we introduce a non-affine extension of the RE within the k-essence framework, focusing on the DBI-type non-canonical Lagrangian \cite{Picon1, Picon2, Babichev1, Sawicki, dg1, dg2, dg3, Vikman,  Chimento, Visser, Scherrer, Padmanabhan,  Mukhanov1, Dutta, Santiago,  Mukohyama,  Das}. K-essence models offer a compelling mechanism for late-time cosmic acceleration without requiring finely tuned potential terms. Originally motivated by the Born-Infeld action—which was developed to resolve the infinite self-energy of the electron \cite{Born, Dirac}—the DBI-type k-essence framework naturally arises in string theory \cite{Callan, Gibbons, Sen} and introduces a modified gravitational metric \cite{gm1, gm2, gm3, gm4}. By incorporating non-affine parametrization, we derive a modified RE for non-geodesic flow, leading to the natural emergence of a dynamical equation of state (EoS) for dark energy. K-essence theory may also address unified dark energy and dust dark matter \cite{Guendelman1}, as well as inflation and dark energy \cite{Guendelman2} in a different context. The Two-Measures Field Theory (TMT) framework demonstrates k-essence dynamics without fine-tuning-- laying the groundwork for scalar field cosmology \cite{Guendelman}. Additionally, recent studies have explored k-essence black-bounce solutions, including magnetically charged configurations with nonlinear electrodynamics \cite{Pereira1},  ghost field extensions preserving energy conditions \cite{Pereira2}, and phantom scalar field ensuring regular spacetime \cite{Pereira3}.

In this letter, we investigate a modified cosmological framework based on a non-affine parametrization of the RE, as induced by a DBI-type k-essence Lagrangian in the emergent metric scenario. In this formulation, we employ a non-affine parameter that introduces additional terms into the RE, thereby modifying the effective expansion dynamics. These terms give rise to a dynamical equation of state (EoS) for dark energy, enabling deviations from the standard $\Lambda CDM$ expansion history.

An anti-damped harmonic oscillator equation for the expansion scalar may be formed from the modified RE. The structure shows that the non-affine parametrization causes a divergence in the expansion scalar, indicating a quantum instability in spacetime development \cite{Panda, Panda1, García, Panda2}.

To test the model, we estimate Bayesian parameters using Hubble, Pantheon+ SN Ia, and BAO observations. Results indicate that non-affine k-essence dynamics modify the Hubble function to allow a higher present-day Hubble constant $H_0$, reducing tension between local and early-universe observations. The variation from the usual EoS accounts for the discrepancy in estimated $H_0$ values at low redshifts, which may be a valid Hubble tension resolution within a single scalar-field framework.

Additionally, an effective repulsive term in the modified RE implies links to emergent gravity and entropic force situations \cite{Padmanabhan1, Verlinde, Verlinde1}, impacting late-time acceleration and quantum gravity models. Observable signatures may also arise in the context of early-universe dynamics, structure formation, and cosmic anisotropies.
\\

It is important to recognize that the geodesic equation, geodesic deviation equation, and Raychaudhuri equation are distinct when analyzing cosmic or gravitational scenarios. The difference is: The geodesic equation uses Christoffel symbols to show a particle or light ray's spacetime path, including curvature. It studies free particle motion and gravitational lensing. The geodesic deviation equation, using the Riemann tensor, examines the relative motion of adjacent geodesics, considering tidal effects, gravitational waves, and geodesic stability. The Raychaudhuri equation emphasizes expansion, shear, and vorticity in geodesic congruence evolution. It is useful in cosmology, singularity theorems, and gravitational focusing. In general, geodesic congruences may be characterized using non-affine parameterizations, not the geodesic equation. Through the Raychaudhuri equation, affine parameterization is a specific example of geodesic congruences. This study is based entirely on the Raychaudhuri equation and does not include the complete framework of the geodesic equation or the geodesic deviation equation.

\section{The K-essence Geometry and the non-affinely parametrized RE}
We explore new geometry emerging from k-essence interaction with the gravitational metric, namely the k-essence geometry. K-essence geometry is derived from non-canonical scalar field theories in which the Lagrangian $\La(X,\phi)$ is not linearly dependent on the kinetic component $X$. It features an emergent metric $\G_{\mu\nu}$, distinct from the background metric $g_{\mu\nu}$, which governs perturbation dynamics. The kinetic term dominates over the potential, allowing for late-time cosmic acceleration without fine-tuning. This framework may generate dark energy at subluminal sound speeds, possibly lowering CMB anisotropies, and induce spontaneous Lorentz invariance breaking, resulting in an analog or emergent spacetime structure that influences cosmic evolution. The effective geometry associated with k-essence is not explicitly given by the action on a flat background; rather, it emerges dynamically through the scalar field’s perturbation behavior, reflecting the field’s intrinsic nonlinear structure and its influence on the causal properties of spacetime. The detail importance, motivation, and summary of applications of the different scenarios of the universe of the k-essence geometry can be found in \cite{Panda1, Panda3}. Based on previous studies \cite{Picon1, Chiba, Picon3, Vikman, Mukhanov1, Babichev1, Das, dg1, dg2, gm1, gm2, gm3} we define our action to be  minimally coupled with gravity:
\ben
S_{k}[\phi,g_{\mu \nu}]=\int d^4x \sqrt{-g} \La(\phi,X)
\label{1}
\een
with $X=\frac{1}{2}g^{\mu \nu} \nabla_{\mu}\phi \nabla_{\nu}\phi$ , $\La(\phi,X)=-V(\phi)F(X)$ is the non-canonical Lagrangian. The corresponding energy-momentum tensor of the k-essence scalar field is:
\ben
T_{\mu \nu}&&=-\frac{2}{\sqrt{-g}}\frac{\delta S_k}{\delta g^{\mu \nu}}=-2\frac{\partial \La}{\partial g^{\mu \nu}}+\La \n \\ &&=-\La_X\nabla_{\mu}\phi \nabla_{\nu}\phi+g_{\mu \nu}\La
\label{2}
\een
where $\La_X=\frac{\partial \La}{\partial X}$, $\La_{X\phi}=\frac{\partial^2 \La}{\partial\phi\partial X}$, $\La_{\phi}=\frac{\partial \La}{\partial \phi}$ and covariant derivative $\nabla_{\mu}$ is defined with respect to the background metric $g_{\mu \nu}$. 

In k-essence geometry, an emergent metric defines the equation of motion (EoM) ({\it viz.} \ref{S2} and \ref{S6}). Here, the emergent metric is not conformally equivalent to the background metric \cite{Bekenstein_1993, gm1, gm2}. The EoM can be simplified under the assumption that the Lagrangian depends only on $X$ and not explicitly on $\phi$ (i.e $\La(\phi,X)\equiv \La(X)$) as,
\ben
\Tilde{G}^{\mu \nu} \nabla_{\mu}\phi \nabla_{\nu}\phi=0,
\label{3}
\een
with the emergent metric $\Tilde{G}_{\mu \nu}=g_{\mu \nu} -\frac{\La_{XX}}{\La_X+2X\La_{XX}}\nabla_{\mu}\phi \nabla_{\nu}\phi$.

In this regard, we should mention that emergent gravity, under the framework of k-essence geometry, posits that gravity and spacetime are not intrinsic entities but emerge from the fundamental dynamics of a scalar field characterized by non-canonical kinetic terms \cite{Babichev1, Vikman}.  Within this paradigm, perturbations of the scalar field propagate not through the background spacetime described by the standard metric $g_{\mu \nu}$, but through an effective geometry governed by the emergent metric $\t{G}_{\mu \nu}$. This emergent metric is explicitly given by after Eq. (\ref{3}) ({\it Appendix} \ref{S5}),
and corresponds to a disformal transformation of the background metric, in the sense introduced by Bekenstein \cite{Bekenstein_1993}. As a result, $\t{G}_{\mu \nu}$ defines a causal structure distinct from $g_{\mu \nu}$, highlighting the emergent and non-fundamental nature of gravity in this context. This means that the world line motions for $\Tilde{G}_{\mu \nu}$ are different from 
$g_{\mu \nu}$ by the term $\frac{\La_{XX}}{\La_X+2X\La_{XX}}\nabla_{\mu}\phi \nabla_{\nu}\phi$ where all the quantities have been defined after Eq. (\ref{2}). Therefore, an observer tied to the metric $g_{\mu\nu}$ will record a different situation. This is the significance of emergent gravity. It will be difficult for an asymptotic (with respect to the scalar particle) observer to ascertain that the difference is owing to a metric that is {\it non-conformal}  to $g_{\mu\nu}$. This indicates that gravity is a macroscopic phenomenon arising from the development of the scalar field, similar to collective behaviors seen in fluids or solids. Collective behaviour in solids and fluids often do not preserve all the microscopic symmetries of individual elements. So $\Tilde{G}_{\mu \nu}$ being non-conformal to $g_{\mu\nu}$ \cite{Bekenstein_1993} is understandable. The notion corresponds with the broader perspective that gravity may stem from more profound microscopic processes rather than being a basic interaction.

We take the form of k-essence DBI kind Lagrangian as \cite{Vikman, Born, Dirac, Padmanabhan}:
\ben
\La(\phi,X)=-V(\phi)\sqrt{1+\frac{2X}{\a(\phi)}}
\label{4}
\een
As we are only interested in the kinetic k-essence model \cite{Scherrer, Padmanabhan} therefore taking $\a(\phi)=-V(\phi)=-1$ simplifies the DBI Lagrangian \cite{Putter, Scherrer} as
\ben
\La(X)=-\sqrt{1-2X}
\label{5}
\een
Although a constant potential simplifies the analysis, our assumption maintains its physical relevance, particularly in the contexts of inflation and late-time acceleration \cite{Babichev, Scherrer, Mukohyama}. The kinetic DBI k-essence Lagrangian extends GR modifications by introducing nonlinear kinetic terms that affect gravity via the scalar field. For this Lagrangian (\ref{5}), the sound speed is $c_s^2 = 1 - 2X$, and the emergent metric ($\Tilde{G}^{\mu \nu}$) is:

\ben
\Tilde{G}_{\mu \nu}=g_{\mu \nu}-\nabla_{\mu}\phi \nabla_{\nu}\phi=g_{\mu \nu}-\partial_{\mu}\phi\partial_{\nu}\phi
\label{6}
\een
$\phi$ is a scalar field.  It should be noted that while Eq. (\ref{6}) or Eq. (\ref{S5}) can be obtained using the gravity coupled with a k-essence scalar field, it does not emerge from Einstein's field equations directly. Instead, it arises as an effective (emergent) metric that governs the propagation of perturbations in the scalar field \cite{Picon1}. For a DBI-type k-essence Lagrangian~ (\ref{4}) or (\ref{5}), the scalar field equation of motion (\ref{S2}), derived from varying the action with respect to $\phi$ (\ref{S1c}), yields a second-order differential equation involving an effective inverse metric $\t{G}_{\mu \nu}$ (\ref{S5}). By analyzing the characteristic surfaces of this equation (e.g., using the eikonal approximation \cite{Babichev1}), one can identify an emergent acoustic metric, which in this special case (DBI-type Lagrangian) reduces to the form provided in Eq. (\ref{6}). Therefore, while Eq. (\ref{6}) is not a solution of the Einstein equations, it is fully consistent with the k-essence framework, which extends general relativity, and reflects the causal structure seen by field perturbations \cite{Babichev1}.

According to Ref. \cite{Padmanabhan}, we consider the energy-momentum tensor in perfect fluid form as:
\ben
T^{\mu}_{\nu}=(\rho + P) u^{\mu}u_{\nu}- P \delta^{\mu}_{\nu},
\label{7}
\een
with $u_{\mu}=\frac{\partial_{\mu} \phi}{\sqrt{2X}}$; $u_{\mu}u^{\mu}=1$; $\rho = 2X\La_{X}-\La=\frac{1}{\sqrt{1-2X}}$; and $P =\La=-\sqrt{1-2X}$.

A unified dark matter and dark energy can be explained via the DBI-type scalar field \cite{Padmanabhan, Scherrer}. By averaging the field properties over different length scales \cite{Padmanabhan}, its behavior transitions naturally, from behaving as pressureless dark matter (dust-like) on small scales, driving structure formation, to exhibiting negative pressure on large scales by smoothening out fluctuations and thereby driving cosmic acceleration. This eliminates the need for separate dark matter and dark energy components while remaining observationally consistent.

Therefore decomposing the k-essence stress-energy tensor into (a) dust-like matter ($P=0$) and (b) a negative-pressure dark energy component ($P=-\rho$)  \cite{Padmanabhan}, yield:
\ben 
\rho =\rho_V+\rho_{DM}, \quad P =P_V+P_{DM} 
\label{8} 
\een
where 
\ben 
&&\rho_{DM}=\frac{2X}{\sqrt{1-2X}}, \quad P_{DM}=0, \n \\ &&\rho_V=\sqrt{1-2X}, \quad P_V=-\rho_V. 
\label{9} 
\een
Here, $\rho_{DM}$ represents cold dark matter, governing early-universe structure formation, while $\rho_V$ corresponds to vacuum energy, whose negative pressure drives cosmic acceleration. The transition from dark matter to dark energy dominance marks a key phase in cosmic evolution, shaping the universe’s expansion history.

As RE is inherently geometric \cite{Kar, Dadhich} in nature, it allows us to extend its use to the emergent metric space influenced by the k-essence scalar field, where the EoM dynamics are modified \cite{Babichev1, Panda, Das}. Therefore, expressing RE in terms of non-affine parameter $\t{s}$ takes the form \cite{Kar, Wald} ({\it Appendix} \ref{S7}--\ref{S12}): 
\ben
\frac{d\t{\theta}}{d\t{s}}+\frac{\t{\theta}^2}{3}=-\t{R}_{\mu \rho}\t{v}^{\mu}\t{v}^{\rho}-2\t{\sigma}^2+2\t{\omega}^2+D_{\mu}\t{A}^{\mu}~~
\label{10}
\een
The given Eq. (\ref{10}) is a scalar identity, ensuring its frame independence and universal applicability in any coordinate system. The last term, $D_{\mu}\t{A}^{\mu}$ , arises specifically due to the non-affine parametrization, accounting for the effects of acceleration in the chosen parametrization. Here, in our case, we take time-like velocity vector as $\t{v}^{\mu} = f(\t{s})\t{u}^{\mu}$, with $f(\t{s})$ being a function depending on the non-affine parameter $\t{s}$, and $\t{u}^{\mu}$ satisfies $\t{u}^{\mu} \t{u}_{\mu} = 1$ in the proper time emergent metric \cite{Wald, Blau}. Here, we define $D_{\mu} \t{v}^{\mu} = \t{\theta}$ as the expansion scalar of the k-essence geometry, and $\dot{\t{v}}^{\mu} = \t{v}^{\rho} D_{\rho} \t{v}^{\mu} = \kappa \t{v}^{\mu} \equiv \t{A}^{\mu}$ as the acceleration, with $\kappa$ being the non-affine parameter defined as  $\kappa= \frac{\dot{f}}{f}$. The shear $\t{\sigma}_{\mu \nu}~(=\frac{1}{2}(D_{\mu}\t{v}_{\rho}+D_{\rho}\t{v}_{\mu})-\frac{1}{3}\t{h}_{\mu \rho} \t{\theta})$ ({\it Appendix}~\ref{S11}) quantifies shape distortion without volume change, while the vorticity tensor $\t{\omega}_{\mu \rho}~(=\frac{1}{2}(D_{\mu}\t{v}_{\rho}-D_{\rho}\t{v}_{\mu}))$ ({\it Appendix}~\ref{S11}) describes flow rotation. The orthogonal hypersurface is $\t{h}_{\mu \rho} = \t{G}_{\mu \rho} - \frac{\t{v}_{\mu} \t{v}_{\rho}}{\lVert v \rVert}$. Due to non-affinity, velocity normalization is crucial for defining hypersurface orthogonality. For non-affinely parameterized congruences, volume evolution deviates from $\t{\theta} = \frac{1}{\sqrt{\t{h}}} \frac{d}{d\t{s}} \sqrt{\t{h}}$, as $\kappa$ introduces extra contributions, potentially leading to non-trace-free shear. This highlights external influences on the non-geodesic flow. Since the congruence is hypersurface orthogonal, Frobenius' theorem ensures the rotation term vanishes \cite{Hawking2, Poisson}. In Eq.(\ref{10}), the curvature term $\t{R}_{\mu \rho} \t{v}^{\mu} \t{v}^{\rho}$ and shear enhance convergence, whereas rotation and acceleration counteract it. We explore how the k-essence scalar field modifies this effect.

The key distinction between affine and non-affine parametrization lies in $f(\t{s})$ ($\kappa\neq 0$), which induces nonzero acceleration. This modification impacts the modified RE, altering the convergence and divergence of non-geodesic flows. Therefore, the term $f(\t{s})$ is essentially responsible for non-gravitational influences, refining the analysis of congruence dynamics beyond pure curvature effects. It is important to note that in our scenario, Eq.~\eqref{10}, the flow is non-geodesic due to the presence of a non-zero acceleration vector $(\tilde{A}^\mu)$, meaning the worldlines do not follow geodesics.  Within the Raychaudhuri framework, geodesic flow requires $\tilde{A}^\mu = 0$, while $\tilde{A}^\mu \neq 0$ indicates a true deviation from geodesic motion, marking the flow as non-geodesic.\\


\section{K-essence driven RE in flat FLRW metric and study of cosmology through data analysis} We derive the modified RE ({\it Appendix Sec. \ref{sec: Deriv-Mod-RE-FLRW-Background}}) with non-affine parametrization in the flat FLRW background, emphasizing the influence of the k-essence scalar field on cosmic acceleration, structure formation, and departures from standard cosmology.

We choose the time-like velocity vector field in the proper time emergent metric as:
\ben
\t{v}^{\a}=f(t)u^{\a} =(f(t),0,0,0)
\label{11}
\een
where $u^{\mu}=(1,0,0,0)$, is the normalized velocity vector field satisfying the flow of geodesic congruences $u^{\mu}D_{\mu}u^{\nu}=0$, but $\t{v}^{\mu}D_{\mu}\t{v}^{\nu}\neq 0$. Consequently, the relationship between the proper time $\tau$ and coordinate time $t$, which is defined as
\ben
\frac{d\tau}{dt}=f(t)=\sqrt{1-\dot{\phi}^2}.
\label{12}
\een 
Here, we take coordinate time $t$ as {\it the non-affine parameter} in modified geometry. The function $f(t)$, which depends on the kinetic term ($\dot{\phi}^2$) of the scalar field ($\phi$), encodes the deviation from affine parametrization and reflects how the presence of the field alters the passage of proper time experienced by fluid elements. This relation (\ref{12}) will be used throughout to express all quantities in terms of the non-affine parameter ($t$). 

In our framework, the non-affine parametrization arises naturally from a transition between the background FLRW metric and an emergent effective geometry induced by the DBI-type k-essence field. This reflects how the effective fluid associated with the field induces non-geodesic motion via additional forces. Therefore, our approach is dynamically equivalent to GR coupled to a k-essence field, where the non-affine formalism provides a covariant description of how non-canonical scalar dynamics modify the congruences. Reformulating the RE in terms of a non-affine parameter derived from the emergent geometry may offer new insight into field-induced acceleration mechanisms. This procedure with non-canonical kinetics governs causal and expansion properties independently of the background geometry.

The {\it modified RE} in terms of the non-affine parametrization and the emergent metric is expressed as ({\it Appendix \ref{S31}}):
\ben
\frac{\Ddot{a}}{a} =- \frac{1}{6}\sum_{i=1}^{2}\t{\rho}_{i}(1+3\t{\omega}_{i})f^2+G(\dot{\phi},\Ddot{\phi})
\label{13}
\een
where we denote $G=\frac{\Ddot{\phi}^2[1-4\dot{\phi}^2]}{9(1-\dot{\phi}^2)^2}$ and $\t{\omega}_{i}=\frac{\t{P}}{\t{\rho}}$ is the EoS parameter in the emergent metric for different components of the k-essence scalar field. Here, in Eq. (\ref{13}), $\t{\rho}_i$ denotes two distinct density components of k-essence geometry in FLRW background, namely $\t{\rho}_V$ and $\t{\rho}_{DM}$. Where the expression of $\t{\rho}_v$ and $\t{\rho}_{DM}$ is provided in Eq. (\ref{15}). Between these components, $\t{\rho}_V$ acts as dark energy, driving late-time acceleration, while $\t{\rho}_{DM}$ behaves as dust, aiding early universe structure formation. Note that here {\it `tilde'} sign is used for Raychaudhuri-based k-essence geometry throughout the article.

In Eq. (\ref{13}), the modified energy density $\t{\rho}_i$, equation of state $\t{\omega}_i$, and function $f$ depends on $\dot{\phi}^2$, thereby altering the effective energy density. The second term of RHS in Eq. (\ref{13})($G(\dot{\phi},\Ddot{\phi})$) represents a non-gravitational force arising from the non-affine parametrization of the emergent metric. Unlike conventional forces, this force stems from the intrinsic properties of the scalar field and its coupling to spacetime geometry through the emergent structure , thereby altering the expansion dynamics independently of Einstein’s equations. This implies that cosmic acceleration (\ref{13}) arises from scalar-field dynamics eliminating the need for a cosmological constant and highlighting the pivotal role of k-essence as the primary driver of acceleration.

The key effect of non-affine parametrization is the emergence of an apparent force-like term and a nonzero shear component which are absent in affine parametrization. The effective acceleration term mimics modified gravity or an external force, while shear introduces anisotropy impacting cosmic structure formation. Thus, non-affine parametrization not only redefines time but also alters the flow structure, inducing force and shear, leading to new physics.

In essence, we explore the non-affine connection of the RE in the k-essence model, adding complexity to dynamics, modifying scalar field behavior and leaving potential observational signatures.

It should be noted that the RE Eq. (\ref{10}) and Einstein's Field Equations (EFE) Eq. (\ref{S23}) are inherently different. The RE is a geometric identity that delineates the focusing of geodesic or non-geodesic flow curves, irrespective of the universe's matter composition \cite{Kar, Dadhich}. It derives the Ricci tensor only by geometric methods, devoid of any direct connection with gravitational theories. Conversely, adopting the Einstein-Hilbert action and applying the variational principle dynamically links the Ricci tensor to matter through the Einstein field equations, which relate spacetime curvature to the energy-momentum tensor. In our study, to explore the universe's dynamical evolution, the Ricci tensor within the FLRW metric must be determined using the Einstein field equations, as it encapsulates the effects of matter and energy. Notably, the Raychaudhuri equation does not fundamentally depend on the Einstein equations—neither explicitly nor implicitly. 

From the EoM (\ref{3}), we get a particular form of the scaling relation \cite{Ganguly} between $\dot{\phi}^2$ and the scale factor $a(t)$ as $\dot{\phi^2}(t)=\frac{C}{C+a(t)^6}$ ({\it Appendix \ref{S19})}, $C$ is an integration constant. But {\it for the purpose of data analysis}, we take generalization of the EoM and maintaining the restriction imposed on $\dot{\phi}^{2}$ as $0<\dot{\phi}^{2}<1$, the scaling equation as:
\ben
\dot{\phi^2}(t)=\frac{C}{C+a^n};~
\frac{\Ddot{\phi}}{(1-\dot{\phi}^2)}=-\frac{nH\dot{\phi}}{2}
\label{14}
\een
where $n$ is treated as a free parameter to achieve a better fit with observational data. With this generalized solution, we proceed to perform model fitting using the latest observational datasets, including the \textit{PANTHEON+SHOES} data \cite{Brout}, \textit{Hubble} data \cite{Abbott, Alam, Fotios,Metin, Aubourg, Julian, Betoule, Beutler1, Beutler2, Gaztanaga, Blake, Xu, Samushia} and \textit{BAO} data \cite{Alam1, Dawson, Desi, Levi, Hussain}.

By using the transformation between emergent stress-energy tensor ($\t{T}_{\mu \rho}$) and the background stress-energy tensor ($T_{\mu\rho}$)({\it Appendix} \ref{S24}), we can express the energy density and pressure of the different components (\ref{9}) of the k-essence geometry corresponding to the emergent metric as:
\ben
&&\t{\rho}_{DM}=\frac{\sqrt{C+a^n}}{a^{\frac{3n}{2}}} \quad , \quad \t{P}_{DM}=0 \n \\ && \t{\rho}_V=\frac{\sqrt{C+a^n}}{a^{\frac{n}{2}}} \quad , \quad \t{P}_{V}=-\frac{a^{\frac{n}{2}}}{\sqrt{C+a^n}}.
\label{15}
\een
Therefore, the EoS parameter in the modified geometry can be expressed as:
\ben
\t{\omega}_{DM}=\frac{\t{P}_{DM}}{\t{\rho}_{DM}}=0 \quad ; \quad \t{\omega}_V=\frac{\t{P}_V}{\t{\rho}_V}=-\frac{a^n}{C+a^n}.
\label{16}
\een
The EoS parameters emerge inherently as a result of the non-affine connection. Therefore, the decomposition of the DBI-type k-essence energy density into pressure-less dark matter and a cosmological constant-like dark energy (\ref{9}) is shown to be frame-dependent. Upon moving to the modified geometry, the dark energy component (\ref{15}) exhibits the similar type of pressure-density relationship of a Chaplygin gas($\t{P}_V=-\frac{1}{\t{\rho}_V}$) \cite{Kamenshchik, Bento}, while the matter sector remains pressure-less. Nevertheless, Chaplygin gas is not our primary concern; it emerges automatically due to the change of frame.

In terms of the above generalization, we can express the {\it modified RE} as ({\it Appendix \ref{S34}}):

\ben
\frac{\Ddot{a}}{a}=-H^2\Big[\frac{1}{2}\sum_{n=1}^{2}(\Omega_i(1+3\t{\omega}_i))+ \frac{n^2}{36}(\t{w}_V+1)(4\t{w}_V+3)\Big] 
\label{17}
\een
where we take $\Omega_{i}=\frac{\t{\rho}_i}{3\t{H}^2}=\frac{\rho_i}{3H^2}$ as dimensionless density parameters.

Another form of {\it modified RE} in terms of redshift distance ($z$) yields ({\it Appendix \ref{S36}}) :
\ben
\frac{dH}{dz}=&&\frac{H}{1+z}\Big(\frac{3}{2}(1+\Omega_V\t{w}_V)+ \frac{n^2}{36}(\t{w}_V+1)(4\t{w}_V+3)\Big)\nonumber\\
\label{18}
\een
We find a differential equation for the rate of change of the dimensionless density parameter ($\Omega_V$) of the dark energy sector with respect to redshift as:
\ben
\frac{d\Omega_V}{dz}=-\frac{\Omega_V}{(1+z)}\Big(\frac{n}{2}(1+\t{w}_V)+\frac{2(1+z)}{H}\frac{dH}{dz}\Big)
\label{18a}
\een 
To constrain the cosmological parameters, we define the dimensionless luminosity distance as $d_l(z)=(1+z)c\int_0^z \frac{dz}{H(z)}$ \cite{Weinberg, Dodelson} and compute its derivative with respect to redshift($z$)as:
\ben
\frac{dd_l(z)}{dz}=\frac{d_l(z)}{1+z}+\frac{c(1+z)}{H(z)}.
\label{18b}
\een
By employing Bayesian inference with the No-U-Turn Sampler (NUTS), a variant of Hamiltonian Monte Carlo \cite{Hoffman}, we estimate cosmological parameters from supernova, Hubble, and BAO data. The model includes differential equations (\ref{18}, \ref{18a}, \ref{18b}) for cosmic evolution, parameterized by 
$H_0$, $\Omega_{V0}$ , $\omega_0$ and $n$ . Likelihoods were computed from the distance modulus difference, Hubble and BAO data, weighted by inverse covariance and standard deviation ({\it Appendix} \ref{S44}, \ref{S45}, \ref{S46}). MCMC sampling \cite{Lewis} yielded posterior distributions, with best-fit parameters extracted from posterior means. The chi-squared ($\chi^2$) statistic assessed goodness-of-fit, while confidence contours ($1\sigma$ \& $2\sigma$) are visualized.

\begin{figure}[ht]
            \centering
        \includegraphics[width=0.46\textwidth]{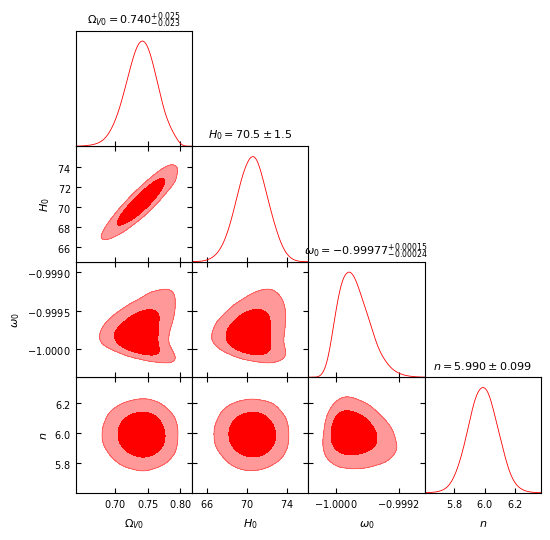}
            \caption{Model fit to Hubble data.}
            \label{FigIa}
        \end{figure}
        
        \begin{figure}[htbp]
            \centering
        \includegraphics[width=0.46\textwidth]{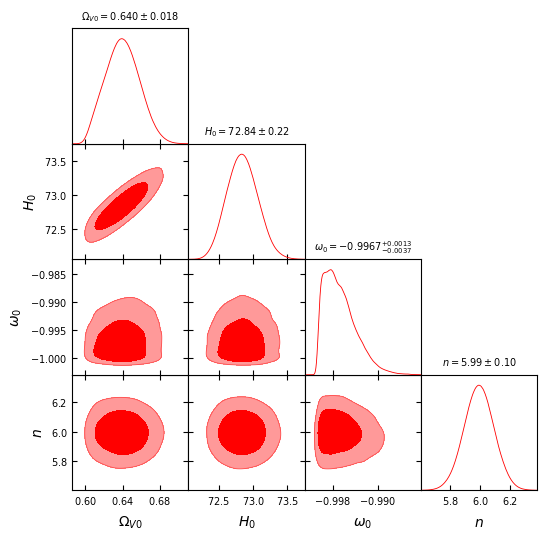}
            \caption{Fit to SN Ia (Pantheon) data.}
            \label{FigIb}
        \end{figure}

        \begin{figure}[htbp]
            \centering
        \includegraphics[width=0.46\textwidth]{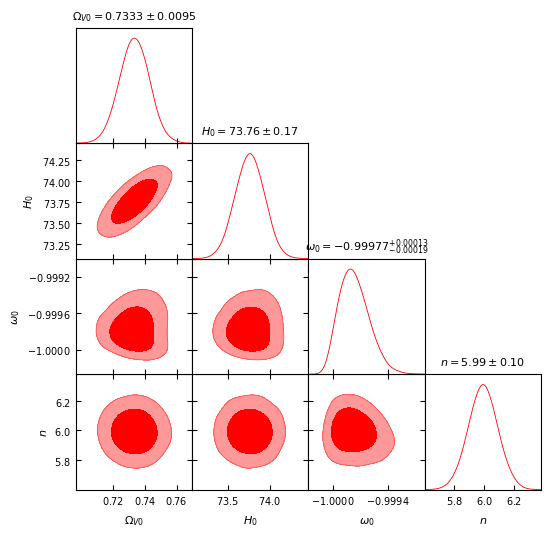}
            \caption{Fit to combined Hubble + Pantheon data.}
            \label{FigIc}
        \end{figure}
        
        \begin{figure}[htbp]
            \centering
        \includegraphics[width=0.46\textwidth]{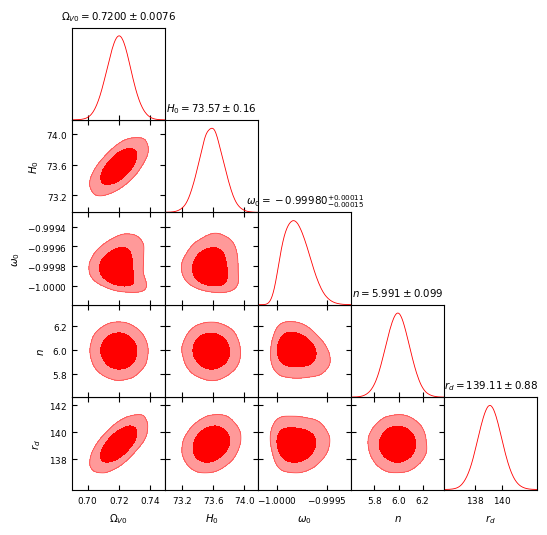}
            \caption{Fit including Hubble, Pantheon, and BAO data, with $r_d$ as an additional parameter.}
            \label{FigId}
        \end{figure}

For all the Figs. (\ref{FigIa}, \ref{FigIb}, \ref{FigIc}, \ref{FigId}), we have to use confidence contours (1$\sigma$, dark; 2$\sigma$, light) for model parameters using different datasets. The fourth plot (\ref{FigId}) incorporates BAO data, introducing $r_d$ to refine the fit.

\begin{table*}[ht]
    \centering
    \resizebox{\textwidth}{!}{ 
    \begin{tabular}{|l|c|c|c|c|c|}
        \hline
        Parameter & Initial Cond. & Hubble Data & PANTHEON+SHOES & Hubble+PANTHEON & Hubble+PANTHEON+BAO \\
        \hline
        $H_0$ & Uniform[50,99] &  $70.5 \pm 1.5$ & $72.84 \pm 0.22$ & $73.76 \pm 0.17$ & $73.57 \pm 0.16$ \\
        \midrule
        $\Omega_{V0}$ & Uniform[0.6,0.8] & $0.740^{+0.025}_{-0.023}$ & $0.640 \pm 0.018 $ & $0.7333 \pm 0.0095$ & $0.72 \pm 0.0076$ \\
        \midrule
        $\omega_0$ & Uniform[-1.2,-0.8] & $-0.99977^{+0.00015}_{-0.00024}$ & $-0.9967^{+0.0013}_{-0.0037}$ & $-0.99977^{+0.00013}_{-0.00019}$ & $-0.99980^{+0.00011}_{-0.00015}$ \\
        \midrule
        $n$ & Gaussian[6,0.1] &  $5.99 \pm 0.099$ & $5.99 \pm 0.10$ & $5.99 \pm 0.10$ & $5.991 \pm 0.099$ \\
        \midrule
        $r_d$ & Uniform[100,300] &  - & - & - & $139.11 \pm 0.88$ \\
        \hline
    \end{tabular}
    }
    \caption{Comparison of cosmological parameters from different data sources.}
    \label{Table1}
\end{table*}

The optimized best-fit parameters are summarized in comprehensive detail in Table \ref{Table1}, providing a detailed comparison of the derived values from different datasets. The dataset used is detailed comprehensively in ({\it Appendix} Tables \ref{TableI},\ref{TableII},\ref{TableIII}). The initial conditions for the parameters are also mentioned in Table \ref{Table1}, where all except $n$ have uniform priors. Instead, $n$ follows a Gaussian prior centered around its expected value to allow slight variations while maintaining theoretical consistency. This soft constraint enables flexibility in parameter adjustment, ensuring a more thorough exploration of parameter space while allowing empirical data to guide the inference. With this, our findings indicate that $n$ remains nearly close to $6$, validating the theoretical prediction.

From the results of the Planck Collaboration (CMB data, 2018), we get $H_0 \approx 67.66\pm 0.42 km/s/Mpc$ \cite{Aghamin} and for the local distance ladder measurement (SHOES) the accepted value is $H_0 \approx 73.2\pm 1.3 km/s/Mpc$ \cite{Riess}. Our analysis refines cosmological parameters by integrating multiple datasets. By analyzing the Pantheon dataset alone or in combination with Hubble data, and subsequently incorporating BAO observations,  results in a almost consistent Hubble constant Table \ref{Table1} with local measurement (SHOES), in contrast with the lowest value of ($70.5 \pm 1.5$ km/s/Mpc) obtained with only Hubble data, thereby highlighting the Hubble tension. This discrepancy suggests new physics beyond $\Lambda CDM$, potentially explained by a dynamical dark energy EoS, early dark energy (EDE), or modified gravity. Additionally, our model predicts a sound horizon ($r_d=139.11\pm0.88$ Mpc), closely aligning with early-universe expansion history as predicted by $\L$CDM model~($r_d= 147.21 \pm 0.23$) \cite{Aghamin} alongside accelerated late-time expansion. This dual effect, preserving early time consistency while allowing for late-time deviations may arise from modifications in cosmic evolution, such as evolving dark energy or altered photon-baryon interactions. Our value of sound horizon is almost fit with the latest observations of BAO measurements by Liu et al. \cite{Liu}. {\it By introducing a time-dependent dark energy parameter, our model may provide a possible resolution to the Hubble tension while maintaining consistency with multiple independent observational probes}. The non-affine extension of the RE in the k-essence framework leads to the {\it emergence of a dynamical EoS parameter for dark energy, which governs a smooth transition from an early matter-like phase to a late-time accelerating phase}. The early time dark energy mimicking dark matter ($\t{\omega}_V\approx 0$) effect increases the total effective matter density, leading to a higher expansion rate of $H(z)$ before recombination. A faster expansion shortens the time available for acoustic oscillation in the primordial plasma, which effectively decreased the inferred sound horizon ($r_d$) when using late-time observation like BAO. This dynamical evolution not only modifies the effective equation governing cosmic acceleration but also naturally leads to a higher inferred local $H_0$, alleviating the observed discrepancy between CMB and local measurements. Unlike models requiring additional new physics, our approach relies solely on {\it scalar field dynamics}, suggesting that {\it the Hubble tension may stem from the oversimplified assumption of a static dark energy component in $\Lambda CDM$ rather than a fundamental inconsistency in cosmological data}.

Additionally, we examine the deceleration parameter with redshift distance. We plot the deceleration parameter ($q$) {\it vs.} redshift ({\it Appendix} \ref{S48}) by using the best-fit parameter obtained from the combined dataset ($PANTHEON+Hubble+BAO$).
\begin{figure}[h] 
    \centering
    \includegraphics[width=0.48\textwidth]{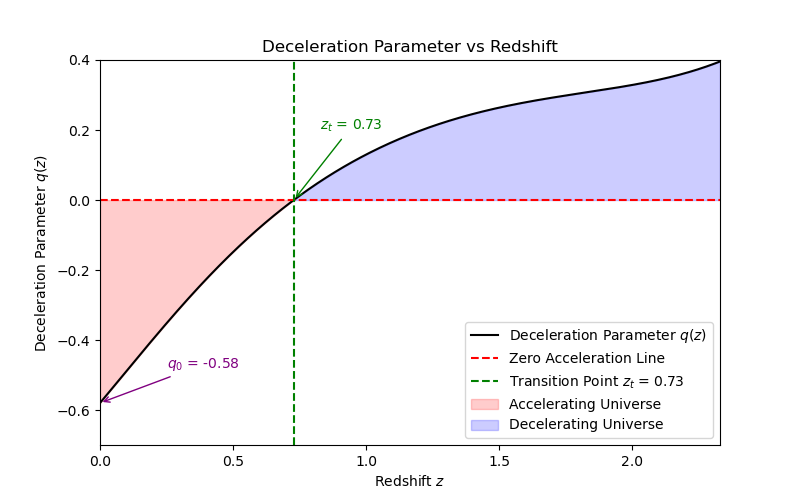}
    \caption{Deceleration parameter ($q$) {\it vs.} Redshift ($z$) graph with parameter scheme $H_0=73.57$, $\Omega_{V0}=0.72$, $\omega_0=- 0.9998$ \& $n=5.991$.}
    \label{Fig2} 
\end{figure}
Our analysis using Pantheon$+$SHOES, BAO, and Hubble data finds $q_0=- 0.58$ \& $z_t=0.73$, indicating a slightly quicker onset of cosmic acceleration compared to $\Lambda CDM$ expectation. While broadly consistent with $\Lambda$CDM ($q_{0}\approx -0.5$), this deviation suggests a {\it modified expansion history where dark energy or an emergent force from non-affine parametrization} influences cosmic evolution earlier. The model naturally accommodates a higher $H_0$ \& $\O_{V0}$ (than $\L$CDM \cite{Aghamin}) by accelerating expansion sooner, which may also offer a compelling resolution to the {\it Hubble tension}.\\

\section{Emergent Oscillatory Dynamics from modified RE}
We redefine $\t{\theta}=3\frac{\dot{\t{F}}}{\t{F}}$ \cite{Kar} and use Eq. (\ref{10}), to express the modified RE as:
\ben
\Ddot{\t{F}}-\kappa\dot{\t{F}}+\frac{1}{3}(\t{R}_{\mu \rho}\t{v}^{\mu}\t{v}^{\rho}-D_{\mu}\dot{\t{v}}^{\mu}+\t{\sigma}^2)\t{F}=0
\label{19}
\een
The evolution of $\t{F}$, related to the expansion of a bundle of flow curves in a non-affine parametrization, follows a damped Hill-type equation \cite{Kar}:
\ben
\Ddot{\t{F}}-\kappa\dot{\t{F}}+\omega^2(t) \t{F}=0.
\label{20} 
\een
The interplay between anti-damping term~($\kappa=\frac{\dot{f}}{f}=\frac{nH\dot{\phi}^2}{2}> 0)$ (\ref{S50}) and time dependent frequency~ ($\omega^2(t)~=\frac{1}{3}(\t{R}_{\mu \rho}\t{v}^{\mu}\t{v}^{\rho}-D_{\mu}\dot{\t{v}}^{\mu}+\t{\sigma}^2))$  may determine the ultimate fate of the expansion scalar, whether it diverges or converges. The presence of the Ricci curvature term $\t{R}_{\mu \nu}\t{v}^{\mu} \t{v}^{\nu}$ in $\omega^2$ accounts for the influence of spacetime curvature on the flow curves, while the shear scalar $\t{\sigma}^2$ introduces anisotropic distortions that tend to amplify the convergence of flow curves. Additionally, the term $D_{\mu}\dot{\t{v}}^{\mu}$ reflects the acceleration of the flow curves (non-geodesic), causing a diverging effect in the presence of non-affinity, due to non-gravitational forces or modified gravity effects in k-essence models. Therefore, the behavior of $\omega^2$ can be determined by these three terms. For $\omega^2 > 0$, flow curves are stabilized by curvature and shear, relevant to expanding universes or perturbation evolution. For $\omega^2<0$, repulsive effects dominate, which may lead to exponential divergence, as seen in inflation, anti-trapped surfaces (inside black hole interiors), or energy condition violations. While the solution for $\omega^2 = 0$ is $\t{F}(t)=c_1 +c_2\int f(t) dt $ where $f(t)=\sqrt{1-\dot{\phi}^2}$ remains positive, which may lead to a runaway divergence under external acceleration.
\begin{figure}[h] 
    \centering
    \includegraphics[width=0.48\textwidth]{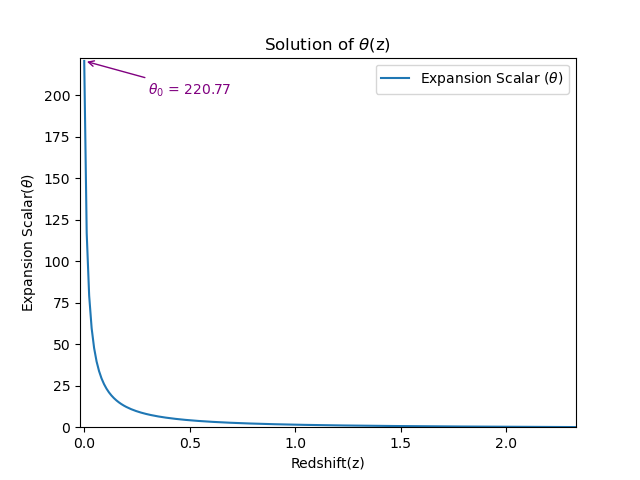}
    \caption{Expansion scalar ($\theta$) {\it vs.} Redshift ($z$) graph with parameter scheme $H_0=73.57$, $\Omega_{V0}=0.72$, $\omega_0=- 0.9998$ \& $n=5.991$.}
    \label{fig3} 
\end{figure}

The term $\kappa$ ({\it Appendix} \ref{S50}) acts as an anti-damping factor governed by the k-essence field, enforces indefinite expansion. Unlike standard geodesic expansion equations \cite{Kar}, non-affine parametrization may alter focusing conditions, it may impact singularity theorems in exotic matter scenarios, thus violating energy conditions. As the sign of $\kappa(t)$ dictates the system's behavior, and in our case $\kappa(t)>0$, the expansion scalar grows indefinitely and may prevent the caustic formation. Let us discuss what happens if the formation of the caustic is prevented: 

In general relativity, caustics occur when geodesic congruences focus and intersect, signaling a local breakdown of the congruence and indicating geodesic incompleteness. However, caustics do not necessarily imply true spacetime singularities, which are global features involving inextendible geodesics, divergent curvature, or loss of predictability.

If caustic formation is avoided, geodesic completeness or possibly a singularity-free universe may be achieved. So, we may say that in our scenarios, caustic prevails indicates that a singularity-free universe via our anti-damping terms. However, in this article, we are not concerned with the study of caustic formation in detail, which is beyond the scope of the present work.

By redefining the Eq. (\ref{20}) in terms of the redshift parameter ({\it Appendix} \ref{S53}) and solving numerically we plot $\t{\theta}=\frac{\dot{\t{F}}}{\t{F}}$ {\it vs.} $z$ (redshift parameter) (Fig. \ref{fig3}) and use the present value of expansion scalar from the best-fit parameters of $Hubble+PANTHEON+BAO$ data. Fig. \ref{fig3} shows a transition from a slower, matter-dominated phase at high redshift to the current accelerated expansion at low redshift. Thereby supporting the avoidance of caustic formation, what Mukohyama et al. \cite{Mukohyama} had demonstrated earlier with DBI scalar field. However, Das et al. \cite{Das} showed that geodesics in DBI-type k-essence models through the affinely connected modified RE can still generate caustics. In k-essence cosmology, non-canonical kinetic terms affect the EoS via the non-affinely connected RE, may avoid singularities, and may indicate {\it quantum effects}.  These features suggest that caustic avoidance in k-essence cosmology is not solely a classical effect but may emerge from the quantum nature of field fluctuations.
However, we have not discussed here the quantum effects thoroughly in our study through Eq. \ref{20}. 

Therefore, our findings may provide new directions for alternative gravity models. By revealing deeper connections, our work lays the foundation for a new class of modified gravity theories in which non-affine parametrization plays a fundamental role in cosmic evolution. Since our model predicts an evolving dark energy component that mimics dark matter at early times, its effects on CMB anisotropies and baryon acoustic oscillations can be further constrained using Planck and other future high-resolution CMB experiments. The predicted repulsive effect emerging from modified RE may have an observational imprint on gravitation waves, which may be tested by LIGO and other gravitational wave detectors. While our model offers a viable resolution to the Hubble tension within the framework of non-affine kinetic k-essence supported by cosmological data from Hubble, Pantheon+ and BAO, its experimental validations through observations such as CMB anisotropies, gravitational waves (LIGO), and large-scale structure surveys lie beyond the scope of this study.\\

{\bf Acknowledgement:}
The authors would like to thank the referee for illuminating suggestions for improving the paper. G.M. would like to extend thanks to all the undergraduate, postgraduate, doctoral students, and all the teachers who significantly enriched him.  S.G. would like to extend his heartfelt appreciation to his wife for her support and continuous encouragement.\\

{\bf Conflicts of interest:} The authors declare no conflicts of interest.\\

{\bf Data availability:} The data used in this study are readily accessible from public sources for validation of our model; however, we did not generate any new datasets for this research.\\

{\bf Declaration of competing interest:}
The authors declare that they have no known competing financial interests or personal relationships that could have appeared to influence the work reported in this paper.\\

{\bf Declaration of generative AI in scientific writing:} The authors state that they do not support the use of AI tools to analyze and extract insights from data as part of the study process.\\

\appendix
\clearpage

\begin{center}
\textbf{\large --- Supplemental Material ---\\ $~$ \\
Non-Affine Extensions of the Raychaudhuri Equation in the K-essence Framework}\\
\medskip
\end{center}
\setcounter{equation}{0}
\makeatletter
\renewcommand{\thesection}{S.\arabic{section}}
\renewcommand{\theequation}{S.\arabic{equation}}

\section{Derivation of k-essence geometry}

The action is
\ben
S_{k}[\phi,g^{\mu \nu}]=\int d^4x \sqrt{-g} \La(\phi,X)
\label{S1}
\een
Here $X=\frac{1}{2}g^{\mu \nu}\nabla_{\mu}\phi\nabla_{\nu}\phi$. Now the first variation of the above action can be expressed as :
\ben
&&\delta S_k[\phi,g^{\mu \nu}]=\int d^4x~ \Big[\sqrt{-g}(\delta \La(\phi,X)) + (\delta \sqrt{-g} )\La(\phi, X)\Big] \n \\ &&=\int d^4x ~\Big[\sqrt{-g}\Big(\frac{\partial \La}{\partial \phi}\delta \phi+ \frac{\partial \La}{\partial X}\delta X\Big)+(\delta \sqrt{-g})\La\Big] \n \\ && = \int d^4x \sqrt{-g} \Big[\frac{\partial \La}{\partial \phi}\delta \phi+\frac{\partial \La}{\partial X} g^{\mu \nu}\nabla_{\mu}\delta \phi\nabla_{\nu}\phi\n \Big]\\ &&+\int d^4x \Big[\frac{\sqrt{-g}}{2}\delta g^{\mu \nu}\frac{\partial \La}{\partial X}\nabla_{\mu}\phi \nabla_{\nu}\phi +(\delta\sqrt{-g})\La\Big]
\label{S1a}
\een

We can express action given in \ref{S1a} as two variations: one with respect to $\phi$ and another with respect to the metric $g^{\mu \nu}$ as: 
\ben
\delta S_{k}[\phi, g^{\mu \nu}]=\delta S_{k}[\phi, g^{\mu \nu}]|_{\phi} + \delta S_{k}[\phi, g^{\mu \nu}]|_{g^{\mu \nu}}
\label{S1b}
\een
Considering only the part of variation depending on $\delta \phi$ denoted by $\delta S_k[\phi,g_{\mu \nu}]|_{\phi}$ and omitting the other terms we can write:
\ben
&&\delta S_k[\phi,g^{\mu \nu}]|_{\phi}= \int d^4x \sqrt{-g}~\Big[\frac{\partial \La}{\partial \phi}\delta \phi  - \nabla_{\mu}(\frac{\partial \La}{\partial X})g^{\mu \nu}\nabla_{\nu}\phi \delta\phi \n \\ &&- \frac{\partial \La}{\partial X}g^{\mu \nu} \nabla_{\mu}\nabla_{\nu}\phi \delta \phi+ \nabla_{\mu}\big(\frac{\partial \La}{\partial X}g^{\mu \nu}\nabla_{\nu}\phi \delta \phi\Big)\Big] \n \\ && = \int d^4x \sqrt{-g}\Big[\La_{\phi}\delta \phi - g^{\mu \nu} \nabla_{\nu}\phi \La_{XX}\nabla_{\mu}X\delta \phi \n \\&&- \La_{X}g^{\mu \nu}\nabla_{\mu}\nabla_{\nu}\phi\delta\phi- g^{\mu \nu}\nabla_{\nu}\phi \La_{\phi X}\nabla_{\mu}\phi\delta \phi\Big]  \n \\ &&=-\int d^4 x \sqrt{-g}\Big[(\La_{XX}\nabla^{\mu}\phi\nabla^{\nu}\phi+\La_{X}g^{\mu \nu})\nabla_{\mu}\nabla_{\nu}\phi \n \\ && +2X\La_{\phi X} -\La_{\phi}\Big]\delta \phi
\label{S1c}
\een

Therefore, the scalar field equation of motion (EoM) is:
\ben
-\frac{1}{\sqrt{-g}}\frac{\delta S_k}{\delta \phi}=\G^{\mu \nu} \nabla_{\mu}\phi \nabla_{\nu}\phi+2X\La_{X\phi}-\La_{\phi}=0
\label{S2}
\een

where $\G_{\mu\nu}$ is the effective emergent metric, which is defined as
\ben
\G^{\mu \nu}=\Big[\La_X g^{\mu \nu}+\La_{XX}\nabla^{\mu}\phi \nabla^{\nu}\phi\Big],
\label{S3}
\een

with $1+\frac{2X\La_{XX}}{\La_X}>0$ and $c_s^2=(1+2X\frac{\La_{XX}}{\La_X})^{-1}$.
The inverse metric is given by \cite{Vikman, gm1, gm2}:
\ben
\G_{\mu \nu}=\frac{\La_X}{c_s}\Big[g_{\mu \nu}-c_s^2\frac{\La_{XX}}{\La_X}\nabla_{\mu}\phi \nabla_{\nu}\phi\Big]
\label{S4}
\een

Using the relation $\Tilde{G}_{\mu \nu}=\frac{c_s}{\La_X}\G_{\mu \nu}$ \cite{gm1, gm2},  we get 
\ben
\Tilde{G}_{\mu \nu}=g_{\mu \nu} -\frac{\La_{XX}}{\La_X+2X\La_{XX}}\nabla_{\mu}\phi \nabla_{\nu}\phi
\label{S5}
\een

Under the assumption of constant potential,$\La(\phi,X)\equiv\La(X)$, the EoM (\ref{S2}) simplifies to:

\ben
\t{G}^{\mu \nu} \nabla_{\mu}\phi \nabla_{\nu}\phi=0
\label{S6}
\een

The other part of variation (\ref{S1b}) with respect to the metric $g^{\mu \nu}$ can be expressed as:
\ben
\delta S_{k}[\phi, g^{\mu \nu}]|_{g^{\mu \nu}}=\int d^4x \sqrt{-g}\delta g^{\mu \nu}\frac{1}{2}\Big[\La_X \nabla_{\mu}\phi\nabla_{\nu}\phi -g_{\mu \nu}\La\Big]\nonumber\\
\label{S6a}
\een

Therefore, the energy-momentum tensor ($T_{\mu \nu}$) is expressed as:
\ben
T_{\mu \nu}=-\frac{2}{\sqrt{-g}}\frac{\delta S_{k}[\phi, g^{\mu \nu}]|_{g^{\mu \nu}}}{\delta g^{\mu \nu}} = \La g_{\mu \nu} -\La_{X}\nabla_{\mu}\phi\nabla_{\nu}\phi
\label{S6b}
\een

\subsection{Derivation of modified Raychaudhuri equation}
The derivation of the RE is inherently geometric, thereby allowing us to extends its use to the modified geometry. In our study, we particularly focus on the time-like velocity vector ($\t{v}^{\mu}$) fields and write \cite{Poisson, Blau}:

\ben
[D_{\mu},D_{\rho}]\t{v}^{\g}=\Tilde{R}^{\g}_{~\a \mu \rho}\t{v}^{\a},
\label{S7}
\een
where the time-like velocity vector field is taken to be $\t{v}^{\mu}=f(\t{s})\t{u}^{\mu}$ where $f(\t{s})$ is some function of our non-affine parameter $\t{s}$ and $\t{u}^{\mu}$ satisfies the orthogonality condition in the emergent metric ($\t{u}^{\mu}\t{u}_{\mu}=1$) \cite{Wald, Blau}. Here $D_{\mu}$ is the covariant derivative in the proper time emergent geometry.
Contracting Eq.(\ref{S7}) with $\t{v}^{\rho}$ and also over the indices $\g$ and $\mu$ we can write, 
\ben
[D_{\mu},D_{\rho}]\t{v}^{\mu}\t{v}^{\rho}=\Tilde{R}_{\a \rho}\t{v}^{\a} \t{v}^{\rho}
\label{S8}
\een
Expanding Eq. (\ref{S8}) we get,
\ben
D_{\mu}\dot{\t{v}}^{\mu}-(D_{\mu}\t{v}^{\rho})(D_{\rho}\t{v}^{\mu})-\frac{d\t{\theta}}{d\t{s}}=\Tilde{R}_{\a \rho}\t{v}^{\a} \t{v}^{\rho}
\label{S9}
\een
where we define $D_{\mu}\t{v}^{\mu}=\t{\theta}$ is the expansion scalar and $\dot{\t{v}}^{\mu}=\t{v}^{\rho}D_{\rho}\t{v}^{\mu}=\kappa \t{v}^{\mu}\equiv \t{A}^{\mu}~(say)$ is the acceleration in the emergent spacetime, with $\kappa$ being the non-affine parameter. This term $\dot{\t{v}}^{\mu}~(=\t{A}^{\mu})$ is non-zero in our study, since we are dealing with the non-affine parameterization of the RE in k-essence geometry.

Now we write the second term of the LHS of Eq. (\ref{S9}) as 
\ben
(D_{\mu}\t{v}^{\rho})(D_{\rho}\t{v}^{\mu})=2\t{\sigma}^2-2\t{\omega}^2+\frac{\t{\theta}^2}{3},
\label{S10}
\een
with  the definition,
\ben
&&\t{\sigma}_{\mu \rho}=\frac{1}{2}(D_{\mu}\t{v}_{\rho}+D_{\rho}\t{v}_{\mu})-\frac{1}{3}\t{h}_{\mu \rho} \t{\theta};\nonumber\\&& \t{\omega}_{\mu \rho}=\frac{1}{2}(D_{\mu}\t{v}_{\rho}-D_{\rho}\t{v}_{\mu}),
\label{S11}
\een

Therefore, we can write the RE in terms of non-affine parameter ($\t{s}$) as:

\ben
\frac{d\t{\theta}}{d\t{s}}+\frac{\t{\theta}^2}{3}=-\t{R}_{\mu \rho}\t{v}^{\mu}\t{v}^{\rho}-2\t{\sigma}^2+2\t{\omega}^2+D_{\mu}\t{A}^{\mu}~~
\label{S12}
\een

\section{Derivation of modified RE in emergent FLRW metric}
\label{sec: Deriv-Mod-RE-FLRW-Background}
\setcounter{equation}{12}

Now, referring to Eq. (\ref{S5}) with the DBI type Lagrangian (\ref{5}), the emergent metric can be expressed as
\ben
d\t{s}^2=(1-\dot{\phi}^2)dt^2-a^2(t)(dr^2+r^2d\theta^2+r^2sin^2\theta d\Phi^2)\nonumber\\
\label{S13}
\een
where $`dot'$ denotes the derivative with respect to coordinate time. Given the homogeneity of the background, we may describe our k-essence scalar field as $\phi\equiv \phi(t)$. To put it another way, we use a homogeneous K-essence scalar field denoted as $\phi(r,t)\equiv \phi(t)$. The dynamical solutions of K-essence scalar fields cause spontaneous Lorentz symmetry violation; hence, the homogenous choice of the field is suitable. According to Eq. (\ref{S13}), the values of $\dot{\phi}^{2}$ must fall between $0$ and $1$, preserving the emergent metric's well-behaved nature.

We take the proper time for a comoving observer as $d\tau=\sqrt{1-\dot{\phi}^2} dt$ and define the emergent metric (\ref{S13}) in the proper time frame to be:
\ben
d\t{s}^2=d\tau^2-a^2(\tau)(dr^2+r^2d\theta^2+r^2sin^2\theta d\Phi^2)
\label{S14}
\een 
We choose the time-like velocity vector field as:
\ben
\t{v}^{\a} =f(t)u^{\a} =(f(t),0,0,0)
\label{S15}
\een
where in the proper time frame, we consider the normalized velocity vector field given by $u^{\mu}=(1,0,0,0)$, which satisfies the geodesic equation $u^{\mu}D_{\mu}u^{\nu}=0$. However, $\t{v}^{\mu}D_{\mu}\t{v}^{\nu}\neq 0$. Here, $D_{\mu}$ is the covariant derivative in proper time frame defined by the connection:
\ben
\Tilde{\Gamma}_{\mu \nu}^{\a}=\frac{\t{G}^{\a \beta}}{2}\big(\partial_{\mu}\t{G}_{\beta \nu}+\partial_{\nu}\t{G}_{\mu \beta}-\partial_{\beta}\t{G}_{\mu \nu}\big)
\label{S16}
\een
Consequently, the relationship between the proper time $\tau$ and coordinate time $t$, which is defined as
\ben
\frac{d\tau}{dt}=f(t)=\sqrt{1-\dot{\phi}^2}.
\label{S17}
\een 
We have chosen the {\it coordinate time $t$ as the non-affine parameter} in the proper time emergent metric (\ref{S14}) through the relation (\ref{S17}). This formulation encapsulates the deviation from affine parametrization, with the function $f(t)$ characterizing the rate at which proper time evolves with respect to coordinate time in the presence of the scalar field $\phi$. In all the subsequent derivations, we consistently express every equation in terms of the non-affine parameter $t$. This approach is justified by the fact that, as long as we adhere to the transformation relations given in (\ref{S17}) and refrain from assuming any specific form of $a(t)$, we are free to employ both proper time and coordinate time metrics interchangeably.

In emergent FLRW metric(\ref{S13}) we can write the EoM as:
\ben
\frac{\Ddot{\phi}}{(1-\dot{\phi}^2)}=-3H\dot{\phi}
\label{S18}
\een
Solving this, we get:
\ben
\dot{\phi^2}(t)=\frac{C}{C+a(t)^6}
\label{S19}
\een
where $C$ is an integration constant.

With that velocity vector field $\t{v}^{\a}$, the acceleration vector is computed as: $\t{A}^{\mu}=\t{v}^{\rho}D_{\rho}\t{v}^{\mu}=\kappa \t{v}^{\mu}$ with $\kappa=\frac{\dot{f}}{f}$ where we define $\kappa=\frac{(\t{v}^{\mu}D_{\mu}\t{v}^{\nu})\t{v}_{\nu}}{\t{v}^{\mu}\t{v}_{\mu}}$ \cite{Wald}. Hence, we get:
\ben
D_{\mu}\t{A}^{\mu}&&=D_{\mu}(\kappa \t{v}^{\mu})= \kappa D_{\mu}\t{v}^{\mu}+\t{v}^{\mu}D_{\mu}\kappa \nonumber\\&&=\kappa \t{\theta}+\t{v}^{\tau}\partial_{\tau} \kappa =\kappa \t{\theta}+\partial_t \kappa
\label{S20}
\een
where we have to use the relation (\ref{S14}) and (\ref{S17}).

Referring to Eqs. (\ref{S11}), (\ref{S15}) and using (\ref{S17}), we obtain the shear:
\ben
\t{\sigma}_{\tau \tau}=\frac{\dot{f}}{f} \quad,  \t{\sigma}_{rr}=\frac{\dot{f}}{3f} \quad , \t{\sigma}_{\theta \theta}=\t{\sigma}_{rr} r^2, 
\t{\sigma}_{\phi \phi}= \t{\sigma}_{\theta \theta} sin^2 \theta\nonumber\\
\label{S21}
\een
In our case, the rotation tensor ($\t{\omega}_{\mu \rho}$) is zero, which follows from Forbenius' theorem \cite{Poisson}. Alternatively, we may set it to $0$ for our isotropic background metric. Since there is no preferred direction in an isotropic background, vorticity ($\t{\omega}^2$) has to vanish. 

In the emergent metric, the curvature term can be expressed in terms of mass-energy density as \cite{Wald, Poisson}: 
\ben
\t{R}_{\mu \rho}\t{v}^{\mu}\t{v}^{\rho}=\Big(\t{T}_{\mu \rho}-\frac{1}{2}\t{T}\t{G}_{\mu \rho}\Big)\t{v}^{\mu}\t{v}^{\rho}
\label{S22}
\een
which we can find from the emergent Einstein's field equation:
\ben
\t{R}_{\mu \rho}-\frac{1}{2}\t{G}_{\mu \rho}\t{R}=\t{T}_{\mu \rho}
\label{S23}
\een
where we consider $\kappa=8\pi G=1$. The corresponding emergent stress-energy tensor ($\t{T}_{\mu \rho}$) and the background stress-energy tensor ($T_{\mu\rho}$) are related by, 
\ben
\t{T}_{\mu \rho}=\frac{\partial x^{\a} }{\partial \t{x}^{\mu} }\frac{\partial x^{\beta} }{ \partial \t{x}^\rho }T_{\a \beta}.
\label{S24}
\een
We can directly express the curvature term  as:
\ben
\t{R}_{\mu \rho}\t{v}^{\mu}\t{v}^{\rho}=\frac{1}{2}(\t{\rho}+3\t{P})f^2
\label{S25}
\een
which stems from our perfect fluid assumption. Here $\t{\rho}$ and $\t{P}$ represent the total energy density and isotropic pressure, respectively, and can be expressed as:
\ben
\t{\rho}=\frac{\rho}{1-\dot{\phi}^2} \quad ; \quad \t{P}=P.
\label{S26}
\een
with respect to the above energy density and pressure, then the equation of state parameter in the emergent metric can be defined as:
\ben
&&\t{\omega}_{DM}=\frac{\t{P}_{DM}}{\t{\rho}_{DM}}=0~;~\t{\omega}_V=\frac{\t{P}_V}{\t{\rho}_V}=-\frac{a^n}{C+a^n}.
\label{S27}
\een
Therefore, we can write:
\ben
\frac{d\t{\theta}}{dt}+\frac{\t{\theta}^2}{3}&& = -\frac{1}{2}\sum_{i=1}^{2}(\t{\rho}_{i}+3\t{P}_{i})f^2-2\kappa^2+\kappa \t{\theta}+\partial_t \kappa \n \\ &&=  -\frac{1}{2}\sum_{i=1}^{2}(\t{\rho}_{i}+3\t{P}_{i})f^2-2\Big(\frac{\dot{f}}{f}\Big)^2+\frac{\dot{f}}{f}  \t{\theta}+\partial_t \Big(\frac{\dot{f}}{f} \Big) \n \\ &&=  -\frac{1}{2}\sum_{i=1}^{2}(\t{\rho}_{i}+3\t{P}_{i})f^2-3\Big(\frac{\dot{f}}{f}\Big)^2+\frac{\dot{f}}{f}\t{\theta}+\frac{\Ddot{f}}{f} \nonumber\\
\label{S28}
\een
where we put all expressions that we have calculated in (\ref{S20}) and (\ref{S21}) . Here $'i'$ stands for the different components of the energy density of the k-essence scalar field denoted as $\tilde{\rho}_V$ (dark energy) and $\tilde{\rho}_{DM}$ (dark matter). 

On the other hand, based on the definition of the scalar expansion ($\t{\theta}$) \cite{Poisson}, we obtain:
\ben
\t{\theta}=D_{\mu}\t{v}^{\mu} =\frac{1}{\sqrt{-\t{G}}}\partial_{\mu}\Big(\sqrt{-\t{G}}\t{v}^{\mu}\Big)=\frac{3\dot{a}}{a}+\frac{\dot{f}}{f} 
\label{S29}
\een
with $\sqrt{-\t{G}}=r^2 sin\theta a^3$ (using (\ref{S14}))  and the derivative of the expansion scalar with respect to the coordinate time $t$ (i.e., non-affine parameter) is written as :
\ben
\frac{d\t{\theta}}{dt}&&=3\big[\frac{\Ddot{a}}{a}-\Big(\frac{\dot{a}}{a}\Big)^2\big]+\frac{\Ddot{f}}{f}-\Big(\frac{\dot{f}}{f}\Big)^2
\label{S30}
\een
The extra term that arises are responsible for the deviation from usual scalar expansion in FLRW metric. The key consequence of the non-affine parametrization is that it generates an apparent force-like term and also introduces a non-zero shear component in a system that would otherwise be free of these effects in affine parametrization.

By using Eq.(\ref{S17}, \ref{S28},\ref{S29}, \ref{S30}) we can write the {\it modified RE} as:
\ben
\frac{\Ddot{a}}{a}&&=\frac{1}{3}\Big[- \frac{1}{2}\sum_{i=1}^{2}(\t{\rho}_{i}+3\t{P}_{i})f^2+\frac{\Ddot{\phi}^2[1-4\dot{\phi}^2]}{3(1-\dot{\phi}^2)^2}\Big] \n \\ &&=- \frac{1}{6}\sum_{i=1}^{2}\t{\rho}_{i}(1+3\t{\omega}_{i})f^2+\frac{\Ddot{\phi}^2[1-4\dot{\phi}^2]}{9(1-\dot{\phi}^2)^2} \n \\ && =- \frac{1}{6}\sum_{i=1}^{2}\t{\rho}_{i}(1+3\t{\omega}_{i})f^2+G(\dot{\phi},\Ddot{\phi})
\label{S31}
\een
where we denote $G=\frac{\Ddot{\phi}^2[1-4\dot{\phi}^2]}{9(1-\dot{\phi}^2)^2}$ and $\t{\omega}_{i}=\frac{\t{P}}{\t{\rho}}$ is the equation of state parameter in the emergent metric for different component of the k-essence scalar field.

Defining the relationship between the scale factor ($a$) to redshift distance ($z$) as \cite{Weinberg, Dodelson}:
\ben
a=\frac{1}{1+z}
\label{S32}
\een

and taking  dimensionless density parameters \cite{Weinberg}:
\ben
\Omega_{i}=\frac{\t{\rho}_i}{3\t{H}^2}=\frac{\rho_i}{3H^2}
\label{S33}
\een

where $'i'$ stands for two different components of the scalar field energy density namely $\t{\Omega}_{V}$ and $\t{\Omega}_{DM}$ corresponding to the dimensionless density parameter of the dark energy and dark matter sector. With respect to the dimensionless density parameter (\ref{S33}), we can write the acceleration term as:
\ben
\frac{\Ddot{a}}{a}= -H^2\Big[\frac{1}{2}\sum_{n=1}^{2}(\Omega_i(1+3\t{\omega}_i))+ \frac{n^2}{36}(\t{w}_V+1)(4\t{w}_V+3)\Big] \nonumber\\
\label{S34}
\een
where we write $G=-\frac{n^2H^2}{36}(\t{\omega}_V+1)(4\t{\omega}_V+3)$.
The Eq. (\ref{S34}) is nothing but {\it another form of modified RE.}

Using Eqs. (\ref{S31}) and (\ref{S34}) we express:  
\ben
\frac{dH}{dz}=\frac{H}{1+z}\Big(\frac{3}{2}(1+\Omega_V\t{w}_V)+   \frac{n^2}{36}(\t{w}_V+1)(4\t{w}_V+3)\Big)\nonumber\\
\label{S36}
\een
where we have used $\dot{z}=-H(1+z)$. This Eq. (\ref{S36}) is also an {\it alternative form of the modified RE in terms of the Hubble parameter ($H$) and redshift ($Z$).}

We find a differential equation for the rate of change of the dimensionless density parameter ($\Omega_V$) of the dark energy sector with respect to redshift as:
\ben
\frac{d\Omega_V}{dz}=-\frac{\Omega_V}{(1+z)}\Big(\frac{n}{2}(1+\t{w}_V)+\frac{2(1+z)}{H}\frac{dH}{dz}\Big)
\label{S37}
\een
with $\t{\omega}_V=-(1-\dot{\phi}^2)$

Here to achieve a better fit with observational data we choose the EoM and corresponding scaling relation to depend on $n$  instead of a fixed value, where $n$ will be used as a free parameter. Thereby changing the Eq. (\ref{S18}) and (\ref{S19}) to:
\ben
\dot{\phi^2}(t)=\frac{C}{C+a^n} \n \\
\frac{\Ddot{\phi}}{(1-\dot{\phi}^2)}=-\frac{nH\dot{\phi}}{2}
\label{S35}
\een

\section{Data Analysis and Model Fitting}
\setcounter{equation}{37}
The \textit{PANTHEON+SHOES} dataset consists of
1701 light curves of 1550 distinct Type Ia supernovae
(SNe Ia) ranging in redshift from z = 0.00122 to 2.2613
\cite{Brout}. The model parameters are to be fitted by comparing the observed and theoretical value of the distance
moduli. The distance moduli can be defined as
\ben
\mu(z,\theta)=5log_{10}(d_l(z))+25
\label{S38}
\een
where $d_l(z)$ is the dimensionless luminosity distance defined as \cite{Weinberg}:
\ben
d_l(z)=(1+z)c\int_0^z \frac{dz}{H(z)}.
\label{S39}
\een
Taking the derivative with respect to z we can obtain a differential equation for $d_l(z)$ as:
\ben
\frac{dd_l(z)}{dz}=\frac{d_l(z)}{1+z}+\frac{c(1+z)}{H(z)}
\label{S40}
\een
where $c$ is the speed of light measured in unit of \textit{km/s}.

The Hubble distance ($D_H$) and the transverse comoving distance ($D_M$),  is measured by the formula:
\ben
&&D_H(z)=\frac{c}{H(z)}.
\label{S41} \\
&&D_M(z)=c(1+z)\int_{0}^{z}\frac{dz'}{H(z')}
\label{S42}
\een
Here,  the sound horizon $r_d$ is the comoving distance that a sound wave could travel in the early universe before the epoch of recombination is defined as:
\ben
r_d=\int_{z_{drag}}^{\infty}\frac{c_s}{H(z)}dz
\label{S43}
\een
Here $z_{drag} \approx 1020$ is the baryon drag epoch, the redshift at which baryons were released from the photon-baryon plasma and $c_s$ is the sound speed of the photon-baryon fluid.
The BAO measurements were also historically summarized by a single quantity
representing the spherically-averaged distance \cite{Eisenstein, Blake2, Percival, Alam1}:
\ben
D_V(z)=[z D_M(z)^2 D_H(z)]^\frac{1}{3}
\label{S44}
\een
This measurement is particularly useful for low-redshift BAO surveys where the separation between transverse and radial measurements is not strong. 

There are two BAO data set we are going to analyze. The first one contains 8 data points from Sloan Digital Sky Survey (SDSS) \cite{Alam1, Dawson, Hussain} called SDSSBAO and the second set contains 7 data points from Dark Energy Spectroscopic Instrument (DESI) \cite{Desi, Levi, Hussain} called DESBAO. In the following discussion we call the combined dataset (SDSSBAO and DESBAO) as BAO data. These data sets are comprehensively tabulated in Table \ref{TableI} and Table \ref{TableII}. 
\begin{table*}[ht]
    \centering
    \renewcommand{\arraystretch}{1.4}
    \begin{tabular}{|c|c|c|c|c|c|c|c|c|}
        \toprule
        $\t{z}$ & \textbf{0.15} & \textbf{0.38} & \textbf{0.51} & \textbf{0.70} & \textbf{0.85} & \textbf{1.48} & \textbf{2.33} & \textbf{2.33} \\
        \midrule
        $D_V(\tilde{z})/r_d$ & 4.47 $\pm$ 0.17 & & & & $18.33^{+0.57}_{-0.62}$ &  &  & \\ \midrule
        $D_M(\tilde{z})/r_d$ & & 10.23 $\pm$ 0.17 & 13.36 $\pm$ 0.21 & 17.86 $\pm$ 0.33 & & 30.69 $\pm$ 0.80 & 37.6 $\pm$ 1.9 & 37.3 $\pm$ 1.7 \\ \midrule
        $D_H(\tilde{z})/r_d$  & & 25.00 $\pm$ 0.76 & 22.33 $\pm$ 0.58 & 19.33 $\pm$ 0.53 & & 13.26 $\pm$ 0.55 & 8.93 $\pm$ 0.28 & 9.08 $\pm$ 0.34 \\
        \bottomrule
    \end{tabular}
    \caption{SDSSBAO measurements}
    \label{TableI}
\end{table*}

\begin{table*}[ht]
    \centering
    \renewcommand{\arraystretch}{1.4}
    \begin{tabular}{|c|c|c|c|c|c|c|c|}
        \toprule
        $\t{z}$ & \textbf{0.295} & \textbf{0.510} & \textbf{0.706} & \textbf{0.930} & \textbf{1.317} & \textbf{1.491} & \textbf{2.330} \\ \midrule
        $D_V(\tilde{z})/r_d$ & 7.93 $\pm$ 0.15 & & & & & 26.07 $\pm$ 0.67 & \\ \midrule
        $D_M(\tilde{z})/r_d$ & & 13.62 $\pm$ 0.25 & 16.85 $\pm$ 0.32 & 21.71 $\pm$ 0.28 & 27.79 $\pm$ 0.69 &  & 39.71 $\pm$ 0.94 \\ \midrule
        $D_H(\tilde{z})/r_d$ & & 20.98 $\pm$ 0.61 & 20.08 $\pm$ 0.60 & 17.88 $\pm$ 0.35 & 13.82 $\pm$ 0.42 &  &  8.52 $\pm$ 0.17   \\
        \bottomrule
    \end{tabular}
    \caption{DESBAO measurements}
    \label{TableII}
\end{table*}
Hubble dataset is provided in the corresponding table \ref{TableIII}.
Here, we construct four-parameter ({\bf p}=($H_0$, $\Omega_{V0}$, $\omega_0$, $n$)) differential Eqs. (\ref{S36}), (\ref{S37}), (\ref{S40}), the solution of which with the initial condition will be fitted against the available data set of the type Ia supernova data (\textit{PANTHEON+SHOES} data) \cite{Brout}, \textit{Hubble} data \cite{Abbott, Alam, Fotios, Metin, Aubourg, Julian, Betoule, Beutler1, Beutler2, Gaztanaga, Blake, Xu, Samushia} and for  \textit{BAO} dataset \cite{Alam1, Dawson, Desi, Levi, Hussain} we need to add another parameter ($r_d$). Therefore to analyze three datasets together we need to construct a five-parameter given as ({\bf p}=($H_0$, $\Omega_{V0}$, $\omega_0$, $n$, $r_d$)). We use $\chi^2$ statistics to constrain the model parameters to measure the discrepancy between observed data and a theoretical model. It is widely used in statistical hypothesis testing and model fitting to assess how well a model describes the given data. For fitting of the Bayesian model \cite{Hoffman}, $\chi^2$ is often used in likelihood functions: $\textit{L} \propto exp(-\frac{1}{2}\chi^2)$. This connects chi-squared minimization to Maximum Likelihood Estimation (MLE), which is useful for Bayesian posterior sampling.

\begin{table}[H]
    \centering
    \begin{tabular}{|c|c|c|c|c|}
        \hline
        z & H(z) & $\sigma_H$ & Method & Ref. \\
        \hline
        0.0708 & 69.0 & $\pm19.68$ & a & \cite{Abbott} \\
        0.09 & 69.0 & $\pm 12.0$ & a & \cite{Alam} \\
        0.12 & 68.6 & $\pm 26.2$ & a & \cite{Abbott} \\
        0.17 & 83.0 & $\pm 8.0$ & a & \cite{Alam} \\
        0.179 & 75.0 & $\pm 4.0$ & a & \cite{Fotios}  \\
        0.199 & 75.0 & $\pm 5.0$ & a &\cite{Fotios} \\
        0.2 & 72.9 & $\pm29.6$ & a & \cite{Abbott} \\
        0.240 & 79.69 & $\pm2.65$ & b & \cite{Gaztanaga} \\
        0.27 & 77.0 & $\pm14.0$ & a & \cite{Alam} \\
        0.28 & 88.8 & $\pm 36.6$ & a & \cite{Abbott} \\
        0.35 & 84.4 & $\pm 7.0$ & b & \cite{Xu} \\
        0.352 & 83.0 & $\pm 14.0$ & a & \cite{Fotios} \\
        0.38 & 81.2 & $\pm 2.2$ & a & \cite{Metin} \\
        0.3802 & 83.0 & $\pm 14.0$  & a & \cite{Aubourg} \\
        0.4 & 95 & $\pm 17.0$ & a &  \cite{Alam} \\
        0.4004 & 77.0 & $\pm 10.2 $ & a & \cite{Aubourg} \\
        0.4247 & 87.1 & $\pm 11.2$ & a & \cite{Aubourg} \\
        0.43 & 86.45 & $\pm 3.68$ & b & \cite{Gaztanaga} \\
        0.44 & 82.6 & $\pm 7.8$ & b & \cite{Blake}  \\
        0.4497 & 92.8 & $\pm12.9$ & a & \cite{Aubourg} \\
        0.47 & 89 & $\pm50$ & a & \cite{Julian} \\
        0.4783 & 80.9 & $\pm 9.0$ & a & \cite{Aubourg} \\
        0.48 & 97.0 & $\pm 62.0$ & a & \cite{Julian} \\
        0.51 & 90.90 & $\pm 2.1$ & a &\cite{Metin} \\
        0.57 & 92.4 & $\pm 4.5$ & b & \cite{Samushia} \\
        0.593 & 104.0 & $\pm13.0$ & a & \cite{Fotios} \\
        0.6 & 87.9 & $\pm6.1$ & b & \cite{Blake} \\
        0.61 & 98.96 & $\pm2.2$ & a & \cite{Metin} \\
        0.68 & 92.0 & $\pm8.0$ & a & \cite{Fotios} \\
        0.73 & 97.3 & $\pm7.0$ & b &\cite{Blake} \\
        0.781 & 105.0 & $\pm12.0$ & a & \cite{Fotios} \\
        0.875 & 125.0 & $\pm17.0$ & a &\cite{Fotios} \\
        0.88 & 90.0 & $\pm40.0$ & a &\cite{Julian} \\
        0.9 & 117.0 & $\pm23.0$ & a & \cite{Alam} \\
        1.037 & 154.0 &$\pm20.0$ & a & \cite{Fotios} \\
        1.3 & 168.0 & $\pm17.0$ & a & \cite{Alam} \\
        1.363 & 160.0 & $\pm33.6$ & a & \cite{Betoule} \\
        1.43 & 177.0 & $\pm18.0$ & a & \cite{Alam} \\
        1.53 & 140.0 & $\pm14.0$ & a & \cite{Alam} \\
        1.75 & 202.0 & $\pm40.0$ & a & \cite{Alam} \\
        1.965 & 186.5 & $\pm50.4$ & a & \cite{Betoule} \\
        2.34 & 222.0 & $\pm 7.0$ & b & \cite{Beutler1} \\
        2.36 & 226.0 & $\pm8.0$ & b & \cite{Beutler2} \\
        \hline
    \end{tabular}
    \caption{Here the unit of $H(z)$ is $km s^{-1} Mpc^{-1}$ 'a' quoted in this table means the $H(z)$ value is deduced from cosmic chronological method/differential age method whereas 'b' corresponds to that obtained from BAO data and the corresponding reference from where the data are collected is mentioned in the References}
    \label{TableIII}
\end{table}

For the model with model parameters ($\bf{p}$) we compute the $\chi^2$ function for \textit{Hubble} dataset given in Table I as \cite{Hogg} :
\ben
\chi_{H}^2=\sum_{i=1}^{43}\Big(\frac{H^{th}(z_i,{\bf p})-H^{obs}(z_i)}{\sigma_i}\Big)^2,
\label{S45}
\een

where $H^{th}(z_i,\bf{p})$ stands for the theoretical value that we obtained by solving the differential equations and $H^{obs}(z_i)$ corresponds to the value given in $H(z)$ of the Table \ref{TableIII} data set and $\sigma_i$ is the error/uncertainties in the measurement of $H(z)$  mentioned in the same table.\\

The $\chi^2$ for BAO data can be computed as:
\ben
\chi_{i}^2=\sum_{i=1}^{N}\Big(\frac{X_i^{th}(z,{\bf p}- X_i^{obs}(z_i)}{\sigma_{i}}\Big)^2,
\label{S46}
\een

where $X_{i}^{th}$ is either of $\frac{D_M}{rd}$, $\frac{D_H}{r_d}$ or $\frac{D_V}{r_d}$ calculated theoretically by model fitting. So the total $\chi^2_{t}$ corresponding to \textit{BAO} dataset is obtained as $\chi^2_{t}=\Big(\chi^2_{(\frac{D_M}{r_d})}+\chi^2_{(\frac{D_V}{r_d})}+ \chi^2_{(\frac{D_H}{r_d})}\Big)$.\\

For the \textit{PANTHEON} data set corresponding to SN Ia supernova we follow a different method to obtain $\chi^2$ as we have a covariance matrix ($C$) of dimension $1701\times 1701$ corresponding to the measurement error of the distance modulus ($\mu(z)$) for all 1701 light curves. The expression of $\chi^2$ is \cite{Hogg}:
\ben
\chi_{SN}^2&&= (\mu_{th}(z_i,{\bf p})-\mu_{obs}(z_i))^TC^{-1}(\mu_{th}(z_i,{\bf p})\nonumber\\&&-\mu_{obs}(z_i))
\label{S47}
\een
where $\mu_{th}(z_i,{\bf p})$ is the theoretical value we obtain from Eq. (\ref{S38}) by solving the differential equations listed in Eqs. (\ref{S36}), (\ref{S37}), (\ref{S40}) with initial condition and model parameter ({\bf p}) as constraints and $\mu_{obs}$ is the observed value available in PANTHEON data set \cite{Brout}  and $C^{-1}$ stands for inverse covariance matrix. \\

we also want to examine how the deceleration parameter changes concerning redshift distance. The expression for the deceleration parameter is :
\ben
q&& =-\frac{\Ddot{a}a}{a^2}=-1-\frac{\frac{dH}{dt}}{H^2}=-1+\frac{1+z}{H}\frac{dH}{dz} \n \\ &&=\frac{1}{2}(1+3\Omega_V\t{w}_V)+ \frac{n^2}{36}(\t{w}_V+1)(4\t{w}_V+3)~~ 
\label{S48}
\een
where we have to use Eq. (\ref{S36}) to write the last expression . We now plot the deceleration parameter ($q$) {\it vs.} redshift by using the best fit parameter obtained from the combined dataset ($PANTHEON+Hubble+BAO$).

\section{Emergent Harmonic Oscillator Model}
\setcounter{equation}{47}

By redefine $\t{\theta}=3\frac{\dot{\t{F}}}{\t{F}}$ , we can express the modified RE (\ref{S12}) as:
\ben
\Ddot{\t{F}}-\kappa\dot{\t{F}}+\frac{1}{3}(\t{R}_{\mu \rho}\t{v}^{\mu}\t{v}^{\rho}-D_{\mu}\dot{\t{v}}^{\mu}+\t{\sigma}^2)\t{F}=0 
\label{S49}
\een
We can write $\kappa$ as, 
\ben
\kappa=\frac{\dot{f}}{f}=-\frac{\Ddot{\phi}\dot{\phi}}{(1-\dot{\phi}^2)} =\frac{nH\dot{\phi}^2}{2}.
\label{S50}
\een
where we use \ref{S35} to write the last expression. We also define time dependent frequency  $\omega^2(t)$ as:
\ben
\omega^2(t)=\frac{1}{3}(\t{R}_{\mu \rho}\t{v}^{\mu}\t{v}^{\rho}-D_{\mu}\dot{\t{v}}^{\mu}+\t{\sigma}^2)
\label{S51}
\een
We can express the Eq. (\ref{S49}) in terms of the redshift parameter by using the transformation (from \ref{S22})
\ben
\frac{d}{dt}\equiv -H(1+z)\frac{d}{dz}
\label{S52}
\een
and using Eqs.\ref{S20},\ref{S21},\ref{S25} and \ref{S36}  as:
\ben
&&\frac{d^2F}{dz^2}+\frac{1}{2(1+z)}\frac{dF}{dz}\Big[3(1+\Omega_V\t{\omega}_V) \nonumber\\ &&+\frac{n^2}{18}(\t{\omega}_V+1)(4\t{\omega}_V+3)+2+n(\t{\omega}_V+1)\Big]+\nonumber\\ && \frac{1}{2(1+z)^2}\Big[3\Omega_V\t{\omega}_V+1+\frac{n^2(\t{\omega}_V+1)}{3}\Big]=0
\label{S53}
\een

\end{document}